\documentclass[12pt,aps,prd,preprint,showkeys]{revtex4}
\usepackage[utf8x]{inputenc}
\usepackage{amsmath}
\usepackage{amsfonts}
\usepackage{amssymb}
\usepackage{graphicx}

\usepackage{url}
\usepackage{mathtools}
\usepackage{subfig}
\usepackage{mathrsfs}

\begin{document}

\title{Decay of baryon inhomogeneities in an expanding universe}
\author{Pratik K. Das, Sovan Sau, Abhisek Saha and Soma Sanyal}
%
%\authorrunning{Sovan Sau et al.} % abbreviated author list (for running head)
%

%
\affiliation{University of Hyderabad, Prof. C.R. Rao Road, Hyderabad 500046, India}
%\email{daspratik.mld@gmail.com\inst{1}, sossp.uoh@nic.in\inst{2}}}
              % typeset the title of the contribution

\begin{abstract}

Baryon inhomogeneities can be generated very early in the universe. These inhomogeneities then decay by particle diffusion in an expanding universe. We study the decay of these baryon inhomogeneities in the early universe using the diffusion equation in the Friedmann–Lemaître–Robertson–Walker (FLRW) metric. We have studied the decay starting from the electroweak phase transition. We calculate the interaction cross section of the quarks with the neutrinos, the electrons and the muons and obtain the diffusion coefficients. The diffusion coefficients are temperature dependent. We find that the expansion of the universe causes the inhomogeneities to decay at a faster rate. We find that the baryon inhomogeneities generated at the electroweak epoch have very low amplitudes at the time of the quark hadron phase transition. So unless inhomogeneities are generated with a very high amplitude (greater than $10^5$ times the background density), they will have no effect on the quark hadron phase transition. After the quark hadron phase transition, we include the interaction of the muons with the neutrons and the protons till 100 MeV. We also find that large density inhomogeneities generated during the quark hadron transition with sizes of the order of 1 km must have amplitudes greater than $10^5 $ times the background density to survive upto the nucleosynthesis epoch in an expanding universe.


\end{abstract}
\maketitle

\section{Introduction}

Primordial cosmological fluctuations are an important part of modern cosmology as they link the current Cosmic Microwave Background Radiation (CMBR) data to the early universe. Initially the theory related to fluctuations was developed by Lifshitz \cite{Lifshitz}. Later on significant work has been done by Hawkings \cite{Hawkings} and Bardeen \cite{Bardeen}. In this work, we are interested in the decay of baryon density fluctuations or inhomogeneities generated in the early universe. Baryon density fluctuations can 
be generated in the electroweak phase transition \cite{heckler,megevand,sanchez} as well as the QCD phase transition \cite{fuller, layek1}. There are various defect mediated mechanisms which generate these inhomogeneities at the electroweak scales \cite{electroweak}. Though baryon inhomogeneities have other roles to play in the evolution of the early universe, the primary method of constraining these inhomogeneities is by studying their effects on the Big Bang Nucleosynthesis calculations. The inhomogeneities generated at the QCD scales have a higher chance of surviving to the nucleosynthesis epoch, hence they have been studied more extensively. The baryon inhomogeneities generated in the electroweak epoch, have less chance of surviving till the nucleosynthesis epoch, hence they are often ignored. However it has been shown that baryon over densities generated in the electroweak epoch can definitely affect the quark hadron phase transition \cite{sanyal} and may also survive till the nucleosynthesis epoch \cite{jedamzik}. One of the effect of these baryon over-densities is to delay the quark hadron phase transition in regions of the inhomogeneities. This, along with the fact that large scale inhomogeneities from the electroweak epoch may survive till the nucleosynthesis epoch and affect the abundances of the light elements makes it important to study the diffusion of particles in an inhomogeneity at the electroweak epoch.

The only detailed study  of the evolution of non-linear sub horizon entropy fluctuations between $100$ GeV and $1$ MeV was done in ref. \cite{jedamzik}. The entropy fluctuations that had been considered in that particular study did not necessarily come from baryon inhomogeneities only, the authors were not interested in the source of the entropy fluctuations. They had assumed certain amplitudes and length scales and evolved them with time. The evolution of the inhomogeneity was performed as a succession of pressure equilibrium states, that dissipate or expand due to neutrino heat transport. The neutrino contribution to the heat transport equation was used and after the quark hadron phase transition the neutron and proton diffusion were taken into account.  They concluded that large scale entropy fluctuations at $100$ GeV could survive until the nucleosynthesis epoch and affect the nucleosynthesis calculations. The main difference between their study and our current study is that we look at baryon inhomogeneities generated at the electroweak scale in the diffusive limit taking into consideration both the quark and the lepton diffusion.  Studies have already shown that baryon inhomogeneities can be generated in the electroweak scale.  Generally, they have an amplitude of around $10^4$ times the background baryon density. Since, the baryon number is carried by the quarks at these high temperatures, the baryon inhomogeneities at these scales will  primarily mean an excess of the number density of quarks in a certain region. These quarks diffuse out by colliding with the leptons, which are the electrons, muons and the neutrinos. In this study we focus on obtaining the diffusion coefficient of these quarks as they move through the plasma. The diffusion coefficient is then used in the particle diffusion equation to study the decay of baryon inhomogeneities in an expanding universe.

The diffusion equation has been studied in the early universe mostly in the context of cosmic rays \cite{cosmicrays}. For the case of baryon diffusion it has been studied mostly in the hadronic phase \cite{sasaki,scherrer,sau} when the baryon number is carried by the neutrons and the protons. Here, for the first time, we use it to study, baryon diffusion in the quark gluon plasma phase. Since we are using the diffusion equation pertaining to an expanding universe, we also extend our study to the QCD scales. We do find that the expansion term makes a significant difference in the decay of the inhomogeneities. We find that in a static universe the decay rate is slow, while in an expanding universe, the inhomogeneities decay much faster. In the electroweak case, the particles that we consider are the quarks, muon, electrons and the neutrinos. For the QCD epoch, we have the hadrons which carry the baryon number. Generally, the inhomogeneities which are generated at the QCD epoch have higher amplitude and sizes than those generated in the electroweak epoch \cite{layek2}. Hence we consider these two cases separately.

In this work, we only look at sub horizon fluctuations. We also assume that the size of the fluctuations are larger than the mean free path of the relevant particles. The diffusion equation, is re-written in the Friedmann–Lemaître–Robertson–Walker (FLRW) metric. We neglect baryon number violating processes at the temperatures around $200$ GeV and assume that the total baryon number density is conserved throughout our calculations. All the baryon density fluctuations we consider are assumed to be Gaussian fluctuations.   

The chief motivation of understanding the decay of the baryon inhomogeneities is the different signatures of Inhomogeneous Big Bang Nucleosynthesis (IBBN) that have been predicted \cite{arbey}. An inhomogeneous BBN will result from patches of region where there is a baryon inhomogeneity \cite{applegate}. However, all the study so far has concentrated only on inhomogeneities generated in the quark hadron phase transition. The only study for entropy fluctuations generated in the electroweak epoch has indicated that there will be no significant decay during the period prior to the quark hadron phase transition. We were motivated to see whether this result will hold for baryon inhomogeneities also. We however, find that this result does not hold for baryon inhomogeneities. The inhomogeneities decay rapidly during this period and they decrease by about three orders of magnitude. Since some of the inhomogeneities generated in the electroweak epoch have an amplitude of only $10^4$, these will decay and have no effect on the quark hadron phase transition or the nucleosynthesis epoch. We also study the decay of the inhomogeneities after the quark hadron phase transition. Though these will undergo significant decay, some of them might survive till the nucleosynthesis epoch. The BBN can then be used to constrain the models that generate these fluctuations \cite{constraints}.

In section II, we discuss the amplitude and size of the baryon inhomogeneities that are of interest to us. In section III we discuss the diffusion equation in the FLRW metric, in section IV, we obtain the diffusion coefficient in the quark gluon plasma phase. Since we take into account the scattering of the quarks with electrons, muons and neutrinos, we have divided this section into three subsections. In section V we present the numerical results of the decay of the baryon overdensities between the temperature $200$ GeV and $200$ MeV. This is the period after the electroweak phase transition and before the quark hadron transition. In section VI, we discuss the decay of the inhomogeneities after the quark hadron transition. Inhomogeneities which survive upto the nucleosynthesis period will affect the light element abundances. In section VII, we summarize our work and present our conclusions.

\section{Baryon inhomogeneities in the early universe}
There are several ways in which baryon over densities can be generated in the early universe. Topological defects such as electroweak strings are unstable and generate baryon number when they decay \cite{Dziarmaga,barriola}. These give rise to local baryon density fluctuations. The scale of these fluctuations will be given approximately by the bubble nucleation distance at the electroweak scale. Inhomogeneities may thus be generated by these strings over small length scales ($~10^{-10}$ cms). These inhomogeneities have a large amplitude and consequently diffuse out. Other than the electroweak strings, superconducting strings are also capable of generating baryon inhomogeneities over small lengthscales \cite{brandenburger}. The most detailed study of baryon inhomogeneity generation, their amplitude and size was done in ref. \cite{megevand}. The lengthscales of the inhomogeneities generated were about $10^{-3}$ cm and they had an amplitude of $10^4$ over the background baryon density. We will not be considering any specific model during this epoch, our focus would be to see if these baryon inhomogeneities can survive at least upto the QCD epoch. We have taken Gaussian fluctuations to represent the baryon inhomogeneities. These are characterised by the amplitude ($A$) 
and the full width at half maxima ($\sigma$). We have kept the $A \sim 10^3$ as it is expected that the baryon inhomogeneities at the electroweak scale are not too high. The $\sigma$ is varied over the horizon range.

For the inhomogeneities at the QCD scale we have taken a higher amplitude. This is because there is a greater probability of generating large over densities at the QCD scale. Inhomogeneities can be generated by moving cosmic strings \cite{layek2}, collapsing Z(3) domain walls \cite{atreya} and inhomogeneous nucleation of bubbles in a first order phase transition \cite{madsen}. Inhomogeneities can be spread over an area of one meter and  can have amplitudes of the order of $10^{12} - 10^{13}$ through these various mechanisms. Larger inhomogeneities may also be generated but they will have lower amplitudes. Again we do not focus on the mechanism that will lead to the generation of these baryon inhomogeneities,  we only study the inhomogeneities over different ranges for an approximate amplitude of $10^4$ times the background baryon density.  Diffusion and decay of baryon inhomogeneities in the QCD epoch has been studied before. In a previous work \cite{sau} we have worked out the decay of baryon inhomogeneities using the diffusion equation but ignoring the expansion of the universe. In this work, we redo the multiparticle diffusion that we had done previously in \cite{sau} for the case of an expanding universe. We find that the expanding universe has a significant contribution to the decay of the large scale inhomogeneities. Whereas in the previous work we had considered inhomogeneities with $\sigma$ values much smaller than the horizon size, in this work we consider large scale baryon inhomogeneities and see how they decay in the expanding universe.  
 
\section{The diffusion equation in the FLRW metric}

The diffusion equation in an expanding universe has been obtained in the context of cosmic ray propagation  \cite{cosmicrays}. We briefly describe it here in terms of baryon diffusion in the early universe. The FLRW metric for the 
flat universe is defined by, 
\begin{equation}
ds^2 = c^2 dt^2 - a^2(t) d\vec{r}^2
\end{equation} 
Here $a(t)$ is the scale factor of the expanding universe and $\vec{r}$ is the spatial coordinate. This is the comoving distance in an expanding universe. Consider a region of the universe with an inhomogeneity given by $n(\vec{r},t)$. As time evolves, the particles in the over dense region tend to move towards the lesser dense region to restore equilibrium and a particle flux is generated. In this case, we consider the diffusion to be isotropic. The local observer then sees the particle flux as, 
\begin{equation}
j_k = - D(t) \frac{\partial}{\partial x^k} n(\vec{r},t)
\end{equation}  
The diffusion coefficient $D(t)$ depends on the scattering cross section and the velocity of the particles. Since there are different kinds of particles in this plasma, we are dealing with multi-particle diffusion here. The conservation of current  gives us, 
\begin{equation}
\frac{\partial}{\partial x^\mu}(\sqrt{g} j^{\mu}) = 0
\end{equation}
Here $\sqrt{g} = a^3(t)$ and using the definition of the Hubble parameter as $H(t) = \frac{\dot{a}(t)}{a(t)}$, the diffusion equation can be written as, 
\begin{equation}
\frac{\partial}{\partial t} n(\vec{r},t) + 3 H(t) n(\vec{r},t) - \frac{D(t)}{a^2} \nabla^2 n(\vec{r},t) = 0
\end{equation}  
This is the diffusion equation that we will solve numerically for a time dependent diffusion coefficient. As mentioned before, the diffusion coefficient depends on the scattering cross section of the particles. Hence it is different at different temperatures. Since the scattering cross sections are obtained in terms of temperatures, we use the time - temperature relation in the radiation dominated universe to convert our time to temperature.
\begin{equation}
t = \frac{(0.95 \times 10^{10})^2}{T^2}
\end{equation} 
Here $t$ is in secs and $T$ is in Kelvin. We then solve the diffusion equation in the FLRW metric numerically over the entire range of temperature from $200$ GeV - $200$ MeV. In the next section, we first present the details of calculating the diffusion coefficient in the quark phase.

\section{Diffusion in the electroweak scale}

We start by studying  particle diffusion in the electroweak scale. The inhomogeneities are formed at
$200$ GeV during the electroweak phase transition. So the diffusion  of particle will start around the same time.  During this epoch the most abundant particles are the quarks, electrons, muons and the neutrinos. Out of all these, it is the quarks which carry the baryon number. So the baryon over-densities would predominantly have a larger density of quarks as compared to the background number density. As the quarks diffuse out of the inhomogeneities trying to reach an equilibrium state they will collide with the electrons, muons and the neutrinos. Here we take two cases depending on the mass of the particles. This is because the quarks are lighter than the muons but heavier than the electrons. So as the quarks move through the muons, we have a lighter gas diffusing into a gas of heavier particles but as the quarks diffuse through the electrons and neutrinos, we have a heavier particle diffusing through a lighter gas\cite{landaufm}.
Since we are not going into the detailed transport equation of the particles, we choose a distribution function for the particles of the light gas. 

 In the first case, for the quarks moving through the muons, the diffusion coefficient is given by, 
\begin{equation}
D =\frac{1}{3N} \left\langle \frac{v}{\sigma_t}\right\rangle = \Bigg(\frac{T}{\pi m}\Bigg)^{1/2} \frac{2^{3/2}}{3 \sigma_t} 
\label{eqn:diffcoeff1}
\end{equation}
In the second case, for a heavier particle moving through a lighter fluid,
to obtain the diffusion coefficient, we have to first compute the mobility of the particle in the background fluid. If the velocity of the particle is $\vec{v}$, then the mobility $b$ is related to the external force ($\vec{f}$) by, $\vec{v} = b \vec{f}$ and the diffusion coefficient is given by, 
\begin{equation}
D = b T 
\label{eqn:diffcoeff2}
\end{equation}
We assume the distribution of particles to be Maxwellian, then the mobility of the particles is given by, 
\begin{equation}
b^{-1} = \frac{16\pi}{T}\int\frac{p^2dp}{3h}vp^2\sigma_te^{-E/T} = \frac{16{\sigma_t}m^2t^2}{3{\pi}^2}.
\end{equation}
Here $\sigma_t$ is the scattering cross-section, $m$ is the mass of the particle. Once the scattering cross section is known, substituting it in the expression for $b$ would enable us to obtain the diffusion coefficient $D$.   
Since the scattering cross-sections are temperature dependent, the diffusion coefficient too would be temperature dependent. To find the diffusion coefficient at these temperatures we therefore, obtain the scattering cross section of the quarks with the leptons. In the next subsections, we will calculate the different scattering cross section for the different interactions.

\subsection{Quark-electron scattering}

We start with the motion of quarks through the electron gas. For this we need to find the scattering cross section for the $e^{-} e^{+}\longrightarrow q \bar{q}$ interaction. The differential cross section is given by, 
\begin{equation}
\frac{d\sigma}{d \Omega} = \frac{Q_f^2 \alpha^2}{2 s} \left(\frac{u^2+t^2}{s^2}\right)
\end{equation}
Here $\alpha \sim 10^{-2} $ is the fine structure constant and $Q_f$ is the momentum transfer in this interaction. The variables $u$, $t$ and $s$ are the Mandelstam variables. 
This gives, 
\begin{equation}
\sigma_{t} = \frac{Q_f^2 \alpha^2}{2s} \int \left(\frac{u^2+t^2}{s^2}\right) (1-cos \theta) d\Omega 
\end{equation} 
The total scattering cross section can be obtained after integrating over the solid angle. The numerical value can be obtained once the energy scale of the colliding particles is known. Since we are working around the electroweak scale, the colliding energy of the particles are also in the GeV range. The mobility factor is thus given by, 
\begin{equation}
b^{-1} = \frac{2 \sigma_t m^2}{3 \pi^2} [8T^2(1- e^{-E/T}) - 2E(2E + 4 T)e^{-E/T}]
\end{equation}
The diffusion coefficient can be calculated numerically after obtaining the mobility at various temperatures. 

\subsection{Quark-neutrino scattering}
Neutrinos do not have any charges, they have weak interactions. Though there are different flavors of quarks as well as neutrinos, since we only need order of magnitude estimations, we just consider, 
\begin{equation}
\sigma_t = \frac{G_F^2 \hat{s}}{\pi}
\end{equation}
Here $G_F$ is the Fermi constant given by, $G_F = 1.166 \times 10^{-5} GeV^{-2}$. Numerically, the cross section turns out to be $\sigma_t = 17.2 \times 10^{-42} cm^2 \times \frac{E_{\nu}}{GeV}$ \cite{farland}. Though we are working at very high temperatures in the GeV scale, the value of the diffusion coefficient is difficult to handle numerically with this value of $\sigma_t$. 
For the numerical calculation we therefore rescale the variables suitably to obtain a stable numerical solution. 

\subsection{Quark-muon scattering}
 
In both the previous cases we had a heavier particle moving through a lighter gas of particles, however the scenario changes considerably when we consider the quarks moving through a gas of muons. For the $\mu^{-} \mu^{+}\longrightarrow q \bar{q}$, though the expression for the interaction cross sections are similar to the electrons, but here the quark is the lighter particle which is moving through a heavier gas of particles (the muons). This means that, 
the diffusion coefficient is given by eqn. \ref{eqn:diffcoeff1}, 
\begin{equation}
D = \left(\frac{1}{2\pi}\right)^{\frac{3}{2}} \left(\frac{s^{\frac{1}{2}}}{Q_f \alpha}\right)^2 \left(\frac{1}{mT} \right)^{\frac{1}{2}} [2T (1-e^{-E/T}) - 2E e^{-E/T}] 
\end{equation}
We have the total cross section given by, 
\begin{equation}
\sigma_t = \frac{4 \pi Q_f^2 \alpha^2}{3 s} 
\end{equation}
In all these cases, we determine the diffusion coefficient at different temperatures numerically. Similar to the quark neutrino cross section, even at such high temperatures, the diffusion coefficient are numerically very large quantities hence for all the different cases we need to do some scaling to obtain numerically stable solutions. This we have done by scaling the energy appropriately so that the value of the diffusion coefficient is of reasonable orders of magnitude. While plotting we have plotted only the amplitude at the different length scales depending on the temperature. The amplitude is dimensionless as it is the ratio of the enhanced density to the background density ($\frac{\Delta n_B}{n_B}$).
Thus the rescaling does not affect the change in the overdensity that we are interested in.  
Only, the length scales are changed appropriately to reflect the decrease in the temperature.   

\section{Decay of inhomogeneities in the quark gluon plasma phase}
We now look at baryon inhomogeneities generated during the electroweak phase transition. Since the maximum amplitude of these inhomogeneities is of the order of $10^4$ times the background density, for our simulations we take a low amplitude Gaussian fluctuation. The amplitude is taken to be of the order of $10^3$. We consider large inhomogeneities whose decay will be affected by the expanding universe. The horizon at these temperatures is of the order of $10 mm$ \cite{schwarz}. 
Though the diffusion coefficients can be obtained numerically, the problem is that they vary considerably in their numerical values. This indicates that the particle content in the inhomogeneity would ultimate define how they decay. Though there are multi particles present in the plasma, we do not go for multi particle diffusion as it becomes numerically quite challenging. We look at each of these interactions separately and see how much each of them contributes to the decay of the baryon inhomogeneity. We believe this will give us some idea of how the baryon inhomogeneity decays in this temperature range. 

The other challenge in studying the diffusion of particles over such a large temperature range, is the fact that the horizon will also change considerably. From $200$ GeV where the horizon is of the order of $1$ cm to $200$ MeV where the horizon is of the order of $10$ kms. We divide it into two parts. We evolve the inhomogeneity from $200$ GeV to about $1$ GeV and then again from $1$ GeV to $200$ MeV. Interestingly, we find that the inhomogeneity decays considerably during this period depending on the particle interactions being considered. However, we find that on the log scale, the amplitude   
decays quite rapidly irrespective of the interactions considered. This means that low amplitude inhomogeneities will be completely wiped out before the quark hadron phase transition. 

Let us first look at the decay of the inhomogeneities between $200$ GeV and $1$ GeV. Here the diffusion coefficients differ considerably as they are dependent on the temperature and this is quite a large temperature range. So we look at the decay of the inhomogeneities for different interactions separately. For all the different figures, we have the baryon inhomogeneity ($\frac{\Delta n_B}{n_B}$), on the y axis and the length scale on the x- axis. In figure 1, we see the decay due to the quarks moving through the electrons. The initial fluctuation is taken at $200$ GeV and the final is taken at $17$ GeV. 
\begin{figure}
\includegraphics[width = 60mm,angle = 270 ]{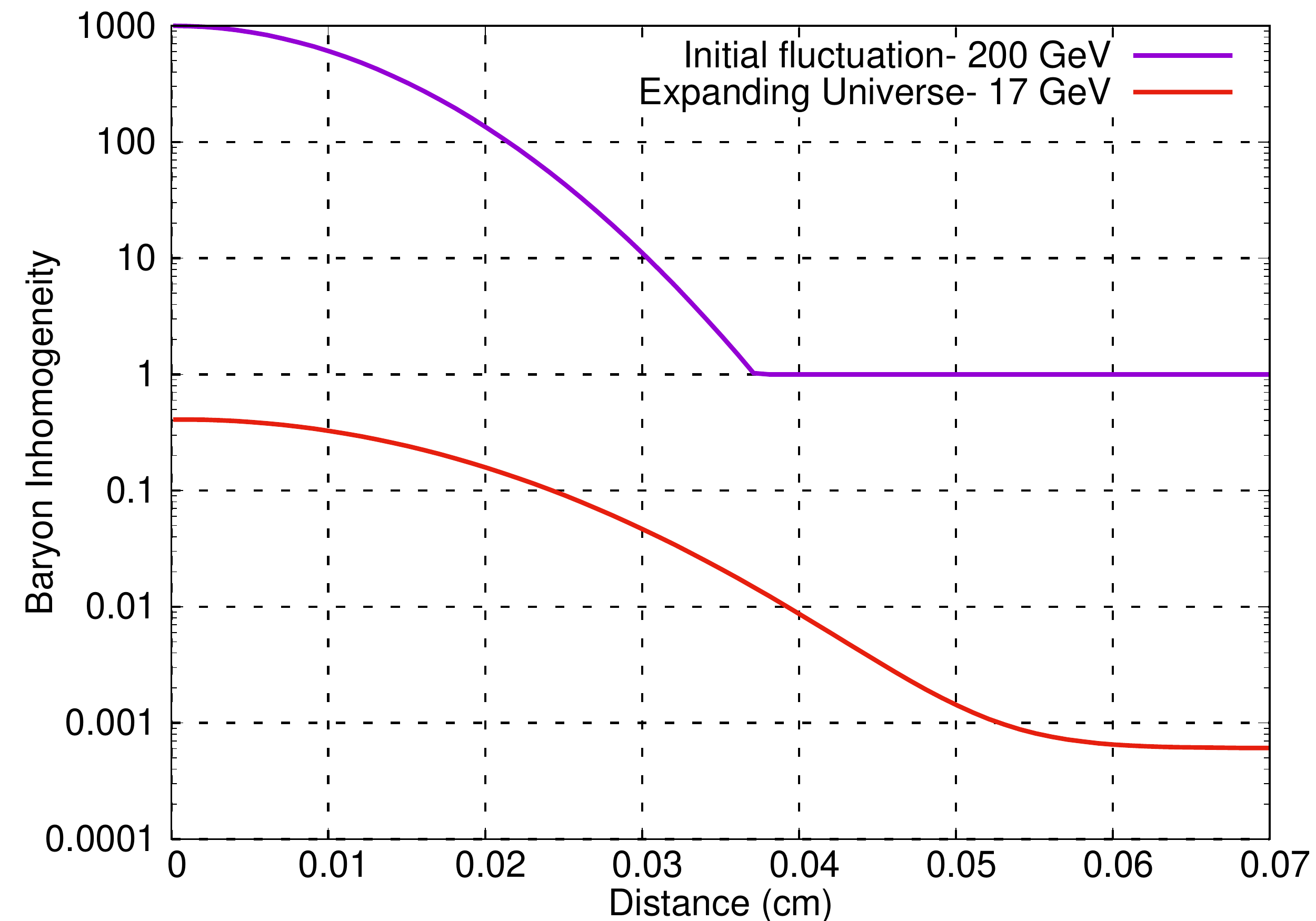}
\caption{The decay of the fluctuation is shown in logscale between $200$ GeV - $17$ GeV as the quarks moves through a sea of electrons. }
\end{figure}
\begin{figure}
\includegraphics[width = 60mm,angle = 270 ]{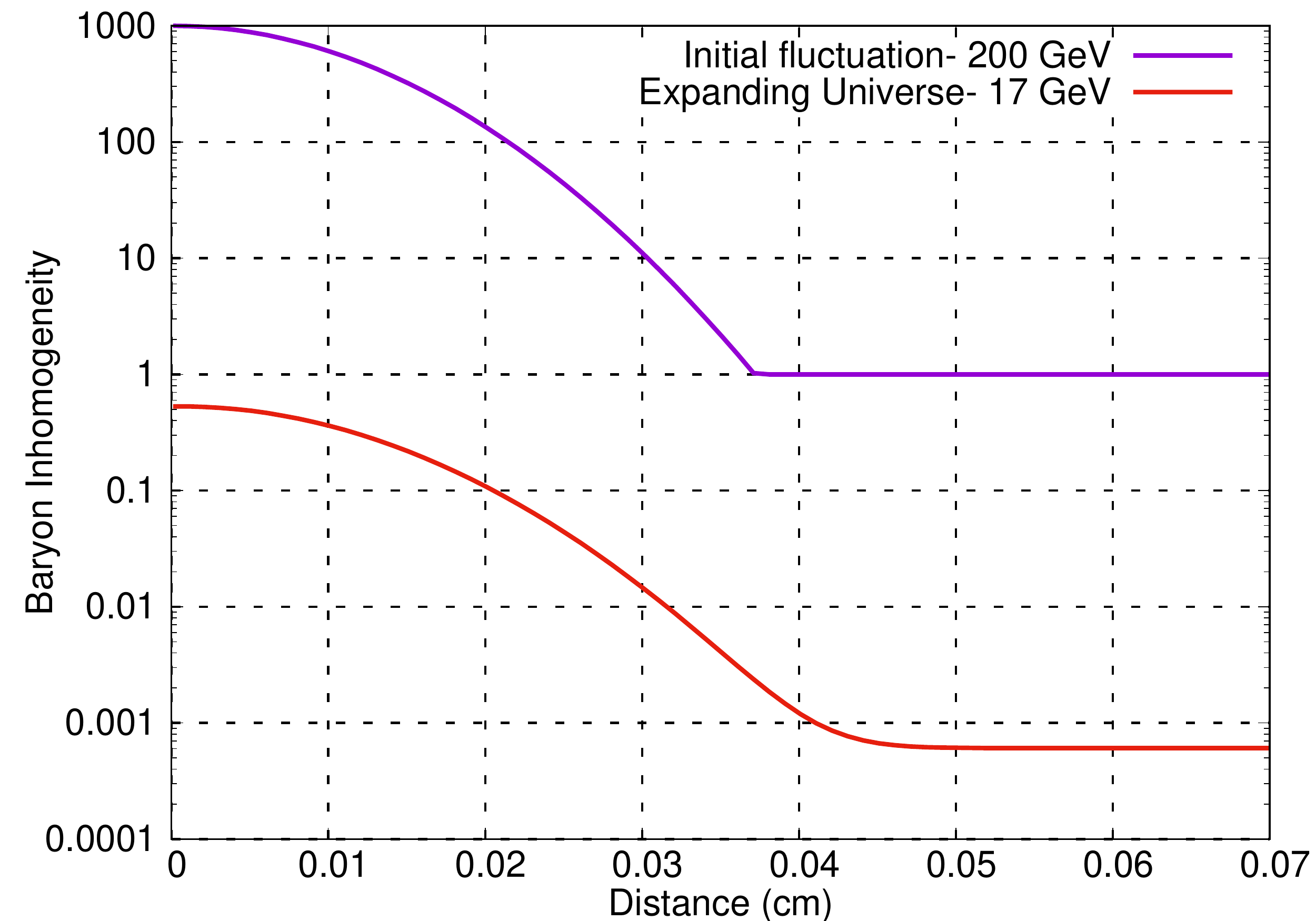}
\caption{The decay of the fluctuation is shown in logscale between $200$ GeV - $17$ GeV as the quarks are predominantly surrounded by neutrinos. }
\end{figure}
As we see the peak of the inhomogeneities goes down by more than three orders of magnitude. The inhomogeneity also spreads out. The decay is similar in the case when the surrounding particles are muons and neutrinos. We have used a different scaling for the neutrinos but as mentioned before the scaling will not affect the relative decay of the amplitude. Though the final amplitude is lower in the case of the neutrinos, the order of magnitude is similar. Since all the graphs have similar decay in orders of magnitude, we have only shown selected graphs.  Fig 1, shows the decay due to the motion of the quarks through the electrons while fig 2 shows the decay due to the motion of the quarks through the neutrinos. Since the plasma at those high temperatures is predominantly dominated by electrons,  our results clearly show that the baryon inhomogeneities decay by about three orders of magnitude in the high temperature GeV range. This is true, even if there are a large number of muons and neutrinos in the plasma. 
 
%\begin{figure}
%\includegraphics[width = 60mm, angle = 270]{N.pdf}
%\includegraphics[width = 60mm,angle = 270 ]{17muqcd.pdf}
%\caption{The decay of the fluctuation is shown in logscale between $200$ GeV - $17$ GeV as the quarks move through muons. }
%\end{figure}

We now look at the decay of the inhomogeneities between the temperatures $1$ GeV to $200$ MeV. We find that the order of magnitude decay is again quite large for the three diffusion coefficients. The individual numbers vary but we plot only the order of magnitude estimates as before. In figure 3, we find that the inhomogeneity has decreased by three orders of magnitude. In the case of fig. 3, the initial fluctuation is at $1$ GeV while the final is plotted at $236$ MeV.    

\begin{figure}
\includegraphics[width = 60mm,angle = 270 ]{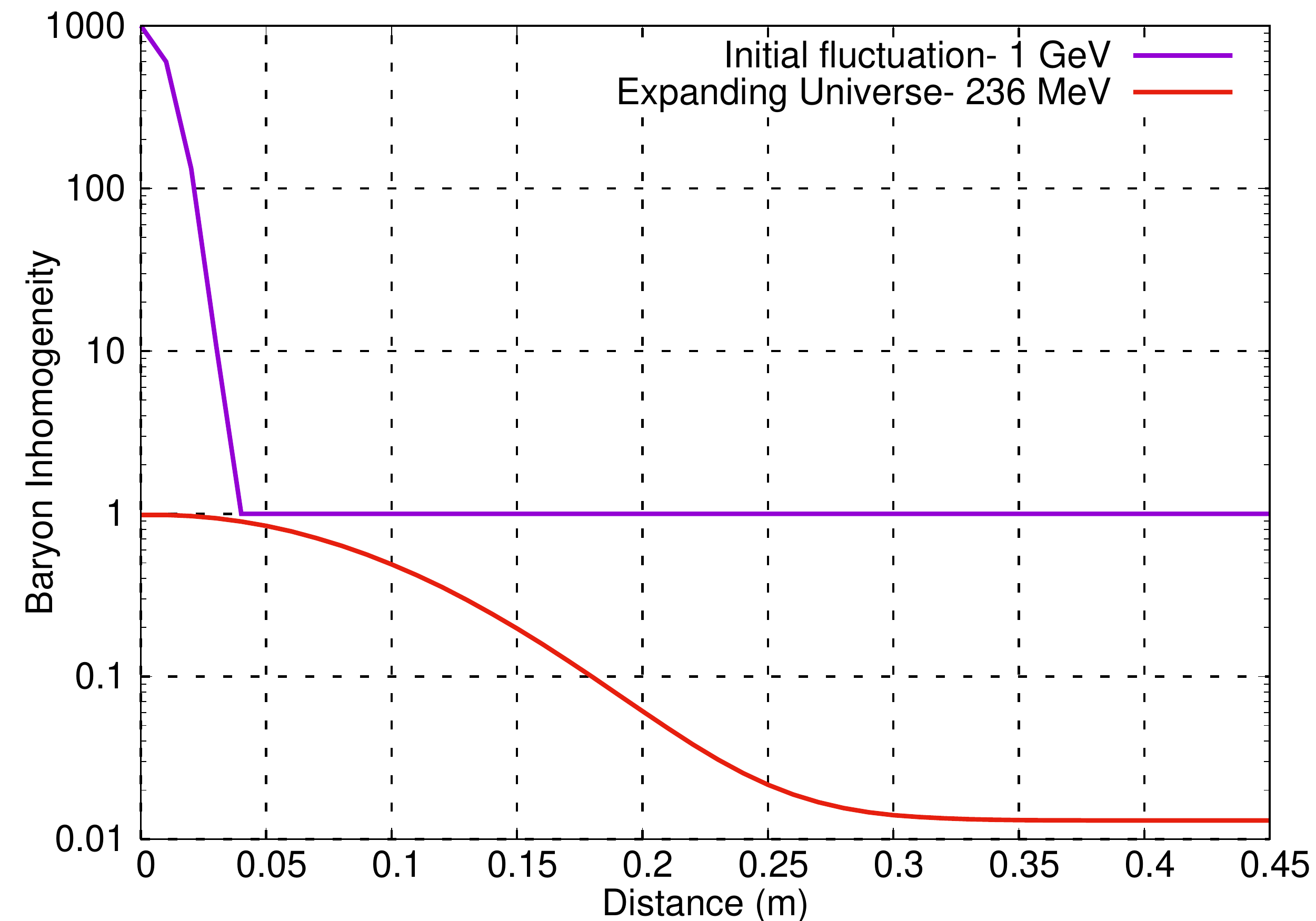}
\caption{The decay of the fluctuation is shown in log scale between $1$ GeV - $236$ MeV for a plasma where a quark is moving through a sea of electrons. }
\end{figure}
For the case of the quarks moving through a large number of muons, the amplitude decay is less than the decay in the case when the particles surrounding the quarks are the electrons. However, the decay is still quite significant.  For the neutrinos, again we have the inhomogeneity decaying by three orders of magnitude. So independent of the particle distribution in the plasma, the amplitude of the inhomogeneities goes down significantly.  In fig 4, we have the quarks moving in a region of muons while in fig 5, the quarks move through the neutrinos. 
\begin{figure}
\includegraphics[width = 60mm,angle = 270 ]{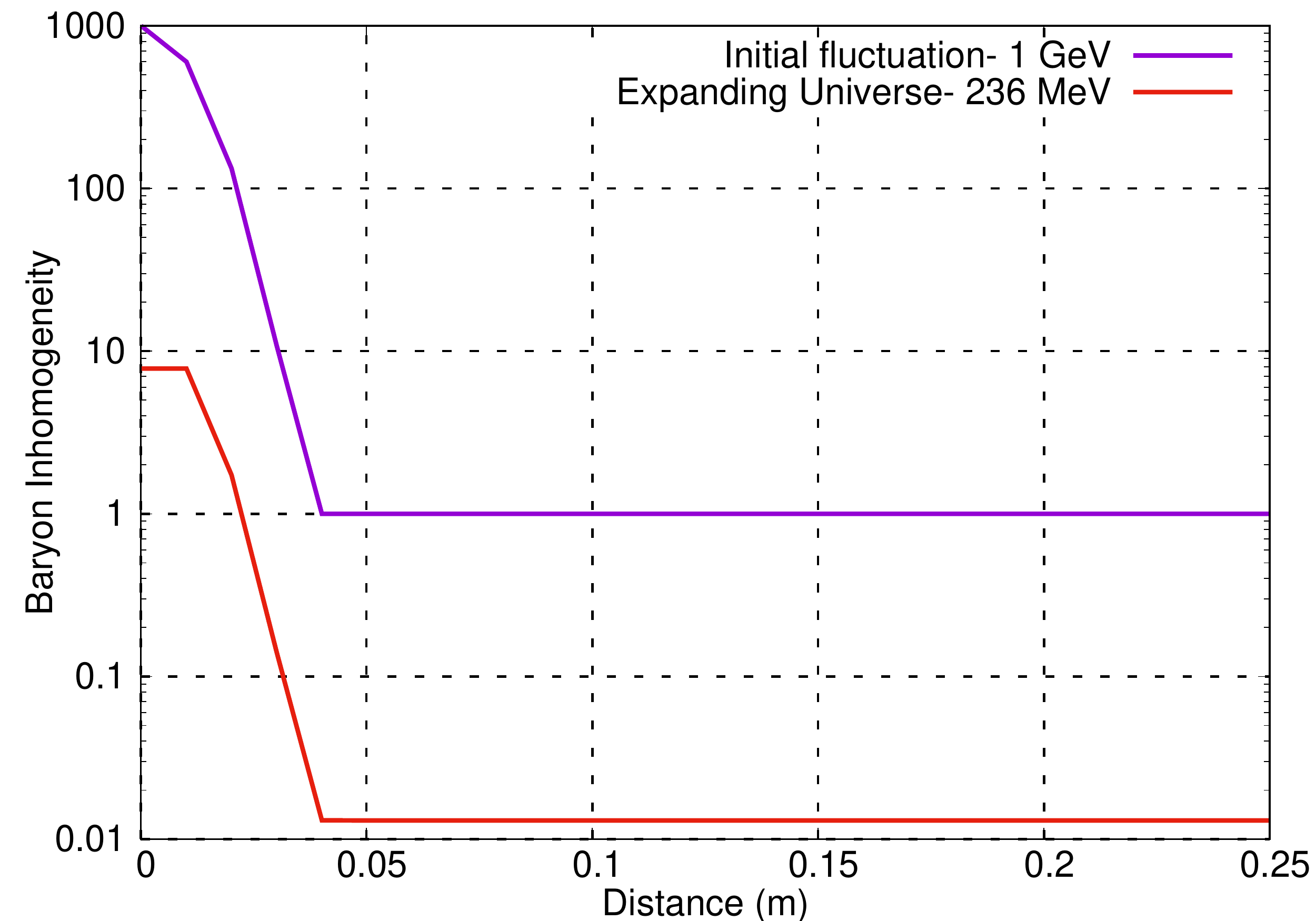}
\caption{The decay of the fluctuation is shown in log scale between $1$ GeV - $236$ MeV for a plasma where a quark is moving through muons. }
\end{figure}
\begin{figure}
\includegraphics[width = 60mm,angle = 270 ]{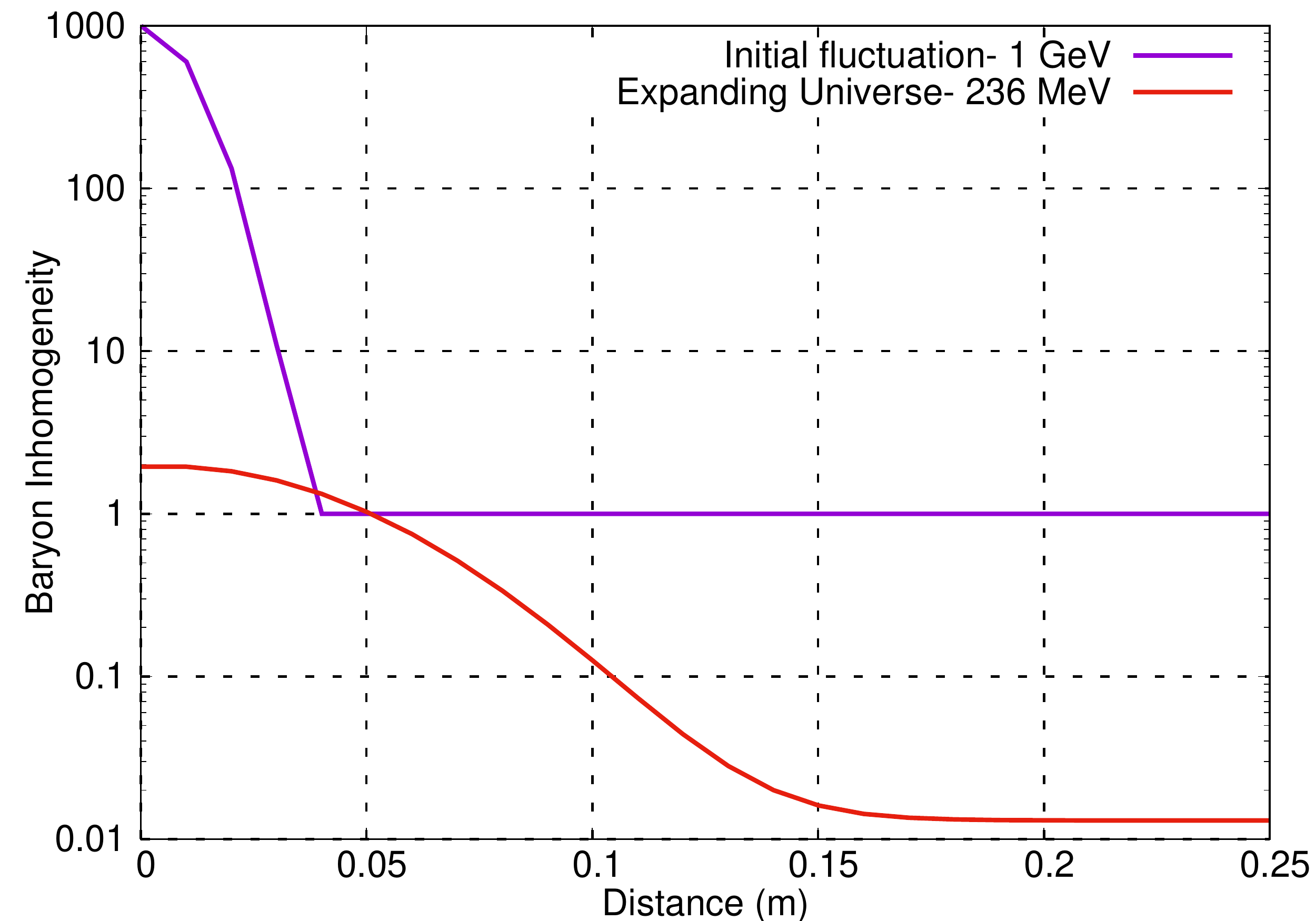}
\caption{The decay of the fluctuation is shown in log scale between $1$ GeV - $236$ MeV for a plasma where a quark is moving through neutrinos. }
\end{figure}
We thus find that the inhomogeneity decays by three orders of magnitude between $200$ GeV to $1$ GeV, and again decays by at least two orders of magnitude between $1$ GeV to $200$ MeV. This means that any inhomogeneity generated at the electroweak epoch needs to have an amplitude greater than $10^5$ times the background density to survive till the quark hadron transition. Thus if we had an inhomogeneity at the electroweak scale with an amplitude less than $10^5$, it would be completely wiped out before the quark hadron transition.   

\section{Decay of inhomogeneities in the hadronic phase}

We have studied the decay of inhomogeneities in the hadronic phase previously \cite{sau}, however in that case we were interested in specific inhomogeneities generated by Z(3) domain walls whose size was much smaller compared to the horizon size. The expansion of the universe was ignored in those cases. We are now interested to see the decay of larger inhomogeneities in the hadronic phase. As mentioned before, inhomogeneities generated during or after the quark hadron phase transition are not only larger in amplitude but they may also be larger in size. Consequently, the decay of these inhomogeneities would be affected by the expanding universe. The plasma during this period consists of the muons, neutrons, protons. electrons and neutrinos. In our previous work, we had shown that the presence of muons enhances the diffusion coefficient of the neutrons/protons, however we had not factored in the expansion of the universe in the previous work. This time we use the diffusion equation for the expanding universe and find that the expansion of the universe causes the inhomogeneities to decay much faster. In the current section, we briefly describe the diffusion coefficient in the hadronic phase and then proceed to present the results of the decay of the inhomogeneities in the hadronic plasma.  

As we have mentioned, the calculation of the diffusion coefficient will depend on which particle is moving through the plasma. The baryon number is carried by the neutrons and the protons, hence here we will be considering the motion of a heavier particle through a lighter gas. The heavier particle is the neutron or the proton, while the lighter gas is a gas of electrons and neutrinos. The muons only play a role till 100 MeV. We thus have to use eqn.\ref{eqn:diffcoeff2} and the scattering cross-section of the neutrons with the electron-positron gas to obtain the diffusion coefficient of the neutrons in the electron positron gas. The scattering cross section is given by, 
\begin{multline}
\frac{d\sigma}{d\Omega}=\frac{\alpha^2\kappa^2q^2}{16M^2E^2sin^4(\theta/2)}\frac{E'}{E}\times
\bigg[1+sin^2(\theta/2)\bigg]
\end{multline}
Here $\theta$ is the scattering angle, while $E$ is the electron energy before the scattering and $E'$ is the electron energy after the scattering. The values of $E$ and $E'$ depend on the temperature of the surrounding plasma. The transport cross section $\sigma_t$, is given by 
\begin{equation}
\sigma_t=\int\frac{d\sigma}{d\Omega}(1-cos\theta)d\Omega
\end{equation}
We can then substitute the scattering cross-section in the transport cross section to obtain, 
\begin{equation}
\sigma_t=3\pi\bigg[\frac{\alpha\kappa}{M}\bigg]^2
\end{equation}
The diffusion coefficient is obtained by substituting the expression for the transport cross-section and we get,
\begin{equation}
 D_{ne} = \frac{M^2}{32 m^3} \frac{1}{\alpha \kappa^2} \frac{e^{1/T}}{Tf(T)}. 
\end{equation}
Here $M$, is neutron mass, $m$ is electron mass and $\kappa = -1.91$ is the anomalous magnetic moment. The temperature in this case is dimensionless as it is scaled by a factor of $m_e c^2$. Finally, the function $f(T)$ is given by, $f(T) = 1 + 3 T + 3 T^2$. 

Similar to the neutron - electron cross section, we can obtain the nucleon-muon scattering cross-section too. We have assumed that the heavy neutron particle is moving through a muon-antimuon gas. The mobility of the neutron is then given by the force on the neutron due to 
the gas. This force is given by the interaction cross section. The differential scattering cross-section is given by,
\begin{multline}
\frac{d\sigma}{d\Omega}=\frac{\alpha^2\kappa^2q^2}{8M^2E^2sin^4(\theta/2)}\frac{1}{1+{2Esin^2(\theta/2)}/M}\times\\
\bigg[\frac{cos^2(\theta/2)}{1-q^2/{4M^2}}\bigg(\frac{q^2}{4M^2}-1\bigg)-2sin^2(\theta/2)\bigg]\\
%\shoveright{\framebox[.55\columnwidth]{\frac{1}{1+{2Esin^2(\theta/2)}/M}\times}}\\
%\framebox[.65\columnwidth]{{4M^2}-1\bigg)-2sin^2(\theta/2)\bigg]}
\end{multline}

%\begin{equation}
%\frac{d\sigma}{d\Omega}=\frac{\alpha^2\kappa^2q^2}{8M^2E^2sin^4(\theta/2)}\frac{1}{1+{2Esin^2(\theta/2)}/M}\bigg[\frac{cos^2(\theta/2)}
% {1-q^2/{4M^2}}\bigg(\frac{q^2}{4M^2}-1\bigg)-2sin^2(\theta/2)\bigg]
%\end{equation}

Here we have assumed that the muon energy and mass are less than the neutron mass. The cross-section  calculation is simplified by this assumption and we can write the cross-section as, 
\begin{equation}
 \frac{d\sigma}{d\Omega} \approx K\frac{\alpha^2\kappa^2}{4M^2}[1+cosec^2(\theta/2)]
\end{equation}
All the constants are replaced by a single constant $K = \frac{1}{2}$. We then substitute the cross section in the diffusion constant. Finally, the diffusion coefficient is given by, 
\begin{equation}
 D_{n\mu}= \frac{M^2}{32m_\mu^3}\frac{1}{\alpha\kappa^2}\frac{e^{1/T'}}{T'f(T')}
\end{equation}
Here $T' = \frac{T}{m_\mu c^2}$. After obtaining both $D_{ne}$ and $D_{n\mu}$, we calculate the total diffusion coefficient of the neutron moving through the plasma.  

Apart from the neutron, we need to find the diffusion coefficient of the proton moving through the electron positron gas too. For proton-electron scattering, the Coulomb force has to be taken into consideration. The scattering cross section for the proton and electron is then given by, 
\begin{equation}
 \frac{d\sigma}{d\Omega} =  \frac{\alpha^2 m_e^2}{4 k^4 sin^4(\theta/2}\bigg[1+ \frac{k^2}{m_e^2}cos^2(\theta/2) \bigg]
\end{equation}
We can obtain the transport cross section from these equations, 
\begin{equation}
\sigma_t = 4 \pi \alpha^2 \bigg[\frac{E_e h}{2 \pi k^2}\bigg]^2 ln (\frac{2}{\theta_0})
\end{equation}
where $\theta_0$ is the minimum scattering angle.
Substituting all the previous equations, we get the diffusion coefficient as, 
\begin{equation}
 D_{pe} = \frac{3 \pi}{8 \alpha^2 ln (\frac{2}{\theta_0})}\bigg[\frac{h}{2 \pi m_e} \bigg] \frac{Te^{1/T}}{f(T)}. 
\end{equation}
Since the muons also constitute a significant part of the plasma till $100$ MeV, we calculate the proton muon cross section too.  The differential cross section is given by, 
\begin{multline}
\frac{d\sigma}{d\Omega}=\frac{\alpha^2}{4E^2sin^4(\theta/2)}\frac{1}{1+{2E sin^2(\theta/2)}/M} \times \\ \bigg[\bigg(1-\frac{\kappa^2q^2}{4M^2}\bigg)cos^2(\theta/2)-\frac{q^2}{2M^2}(1+\kappa)^2sin^2(\theta/2)\bigg]
\end{multline}
We obtain the numerical value of this diffusion coefficient by substituting the constants in the 
transport cross section. Once we have the diffusion coefficients, we numerically solve the diffusion equation in the FLRW metric.

As mentioned before we are considering inhomogeneities whose sizes are in the range of $1$ km. Since the horizon size is around $10$ kms in the hadronic phase, these are large inhomogeneities. We have considered high amplitudes of the order of $10^{14}$ as well as smaller amplitudes, we find that the decay rate does not depend significantly on the amplitudes. However, we find that in an expanding universe the overdensity falls far more rapidly than in an non-expanding universe. We have shown both the cases in figure 6. for comparison. We have checked for the decay separately in the range $200$ MeV - $100$ MeV as the muon is still present in the plasma at these temperatures.  At lower temperatures the muon density in the plasma becomes negligible.

\begin{figure}
\includegraphics[width = 60mm, angle = 270]{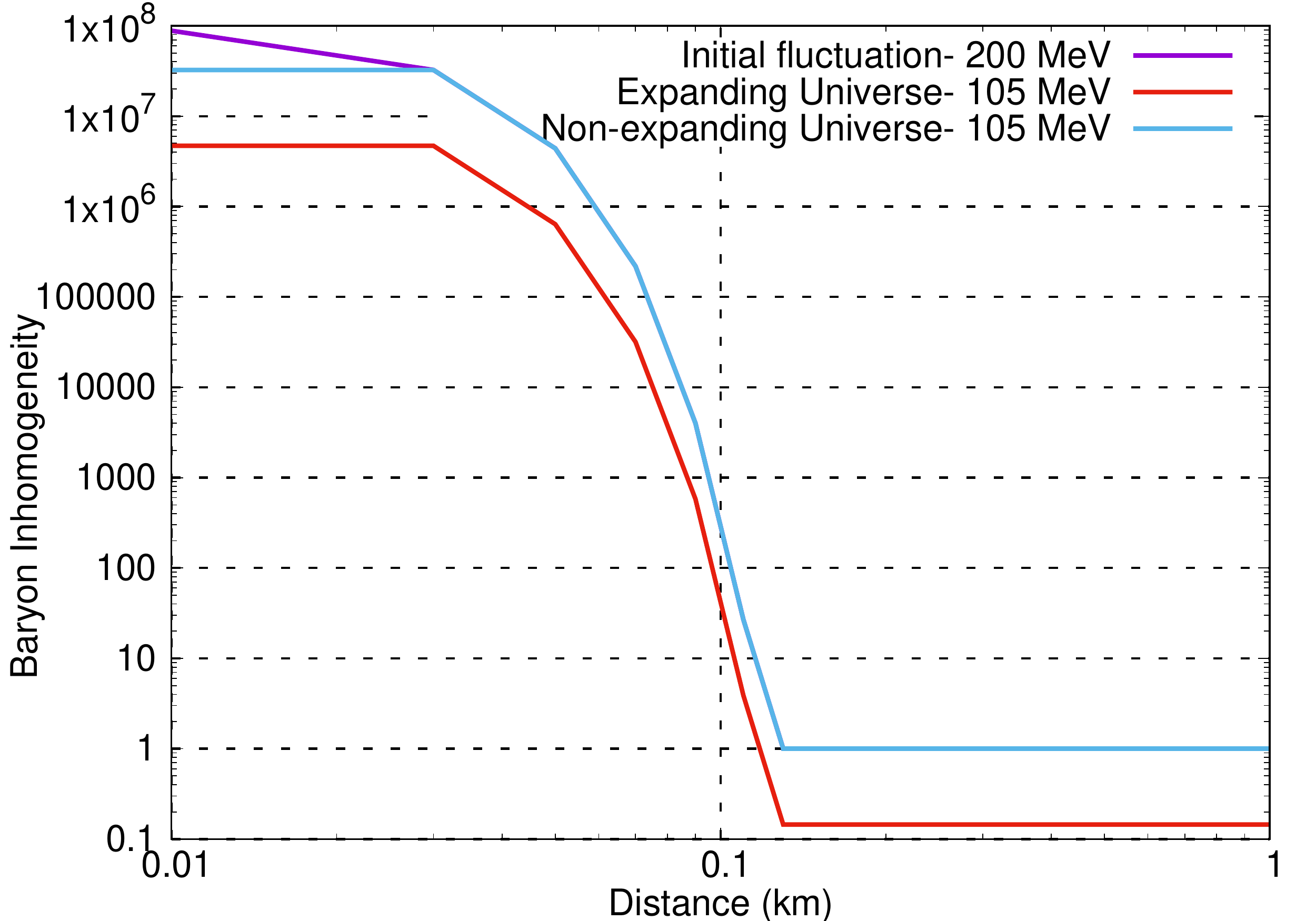}
\caption{The decay of the initial fluctuation is shown in logscale between $200$ MeV - $100$ MeV. }
\end{figure}

It seems that large inhomogeneities do decay significantly in an expanding universe but as long as they have very large amplitude, they may still survive upto the nucleosynthesis epoch. So an inhomogeneity whose amplitude is of the order of $10^8$ will be decreased to an amplitude of the order of $10^7$. Hence inhomogeneities with low amplitudes of the order of $10$ will be wiped out.  Finally, we look at the temperature range from $100$ MeV - $1$ MeV. The muons will be negligible in this epoch but the diffusion coefficients will not change. Figure 7 shows the decay of the inhomogeneities in this epoch.  
\begin{figure}
\includegraphics[width = 60mm, angle = 270]{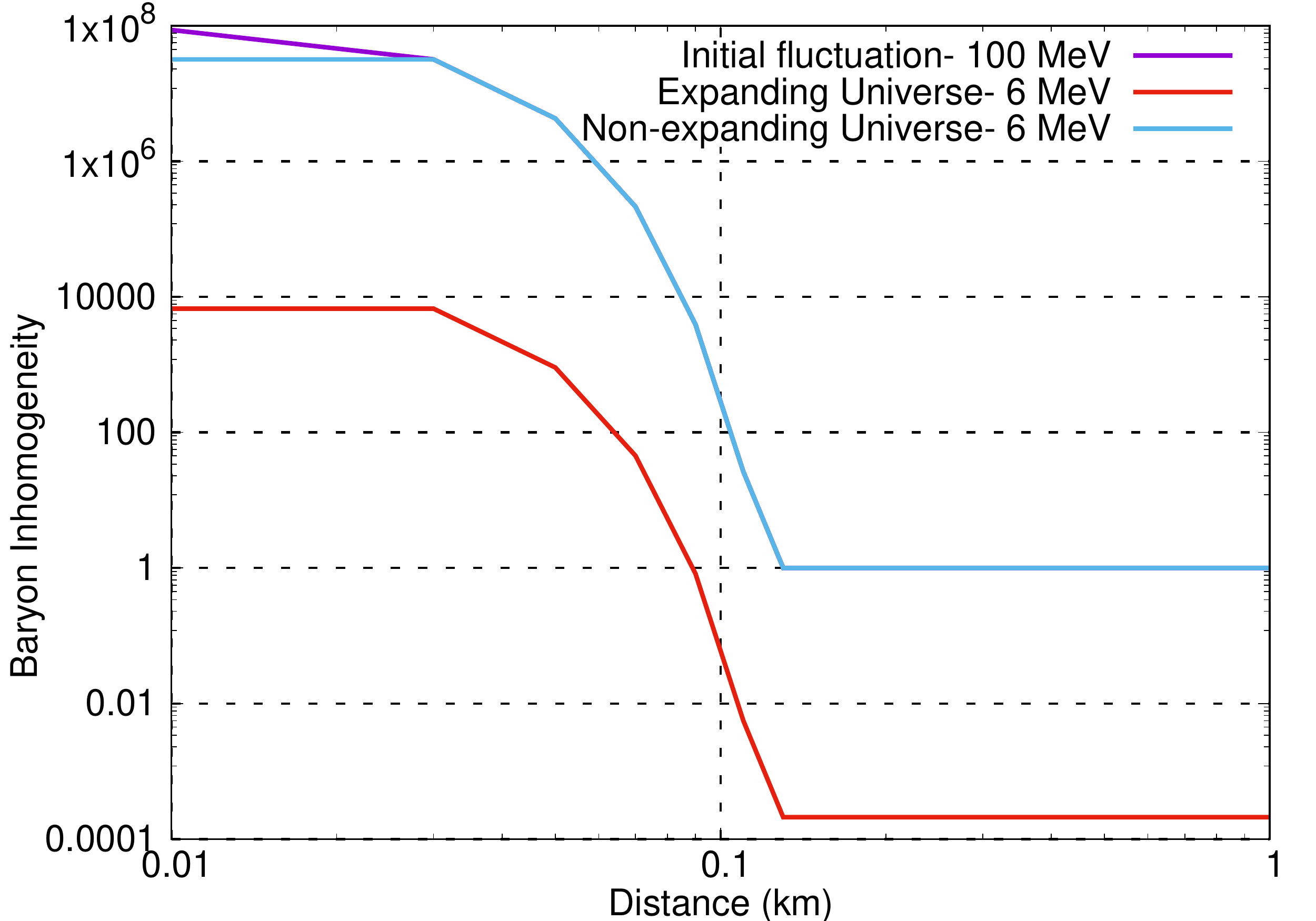}
\caption{The decay of the initial fluctuation is shown in logscale between $100$ MeV - $1$ MeV. }
\end{figure}
We find that the amplitude of the inhomogeneity decreases by an order of $10^4$ in this period. This means that any inhomogeneity with an amplitude less that $10^4$ will be wiped out before the nucleosynthesis epoch. So large baryon inhomogeneities generated during the quark hadron transition must have amplitudes greater than $10^5$ times the background density to survive till the nucleosynthesis epoch.

\section{Summary}
In summary, we have done a detailed study of the decay of the baryon inhomogeneities generated at the electroweak scale.
Baryon inhomogeneities have important consequences in the early universe. If they survive till the quark hadron phase transition they will affect the phase transition dynamics. The quark hadron phase transition is very important in the thermal history of the universe. Moreover, baryon inhomogeneities can also be generated during the quark hadron phase transition. These will have an effect on the Big Bang Nucleosynthesis calculations. Thus the decay of baryon inhomogeneities are important in the early universe. There has been no previous studies of the decay of baryon inhomogeneities in the early universe during the electroweak scale. We have studied the decay of these inhomogeneities in the presence of electrons, muons and neutrinos. The baryon number is carried by the quarks at these high temperatures, so as the inhomogeneity decays, the quarks diffuse through the electrons, muons and neutrinos. The diffusion coefficients for the different particle interactions are calculated. We then use these diffusion coefficients to study the diffusion of the baryon inhomogeneity using the diffusion equation in the FLRW metric.

We have found that baryon inhomogeneities generated in the electroweak epoch should have an amplitude greater than $10^5$ for them to survive till the quark hadron phase transition. This makes it difficult for the baryon inhomogeneities generated in a first order electroweak phase transition to have any effect on the quark hadron epoch. The baryon inhomogeneities have to have a very high amplitude,  they decay substantially during the period between $200$ GeV - $400$ MeV. This is because the diffusion coefficient is temperature dependent. The quark hadron transition occurs around $200$ MeV. We have found that the amplitude of the baryon inhomogeneity decreases to about five orders of magnitude during this period. This means any inhomogeneity with an amplitude of $10^5$ (or less) will be wiped away before the quark hadron phase transition. We therefore conclude that any model which generates inhomogeneities with less than $10^5$ amplitude in the electroweak epoch cannot affect the quark hadron phase transition. They will therefore not contribute to inhomogeneous BBN either.   

Finally in a previous work, we had looked at the decay of baryon inhomogeneities in the QCD epoch for a stationary universe. This would work only for small scale inhomogeneities for which the expansion of the horizon does not matter. 
We have extended that work for an expanding universe where we can work with large baryon inhomogeneities. So we look at the decay of large baryon inhomogeneities in the QCD epoch. We find that the baryon inhomogeneities decrease by $5$-$6$ orders of magnitude. This means that if large baryon inhomogeneities are generated by collapsing domain walls and other topological defects during the quark hadron transition they will survive till the nucleosynthesis epoch. We conclude that  the big bang nucleosynthesis, can thus be used to constrain models which generate large amplitude inhomogeneities in the QCD epoch only.

\begin{center}
 Acknowledgments
\end{center} 
The authors acknowledge discussions with Soumen Nayak and Salil Joshi. A.S is supported by the INSPIRE Fellowship of the Department of Science and Technology (DST) Govt. of India, through Grant no: IF170627.

%
% ---- Bibliography ----
%

\end{document}